\documentstyle[prd,aps,preprint]{revtex}

\def\[{\left [}
\def\]{\right ]}
\def\({\left (}
\def\){\right )}

\newcommand\be{\begin{equation}}
\newcommand\ee{\end{equation}}
\newcommand\bea{\begin{eqnarray}}
\newcommand\eea{\end{eqnarray}}

\newcommand\GeV{\,\mbox{GeV}}

\newcommand\mpl{M_{\rm Pl}}

\def\lsim{\lesssim}

\tighten

\begin{document}
\begin{titlepage}
\pagestyle{empty}
\begin{center}
\tighten

    \hfill  NKPDM-990401 \\ 
\vskip .2in
{\large \bf Particle Physics Inflation Model Constrained from
Astrophysics Observations}\footnote{\tighten{
This work was supported in part by the National Nature Science Foundation of P.R.China} }
\vskip .2in
\vskip .2in
Meng, Xin He 
\vskip .2in
{Department of Physics, Nankai University, Tianjin 300071, P.R.China\\}
{\em Email  mengxh@public1.tpt.tj.cn\\}
and\\
{\em CCAST(World Lab),PO Box 8370, Beijing 100080, P.R.China \\ }
\end{center}

\tighten

\begin{abstract}
The early Universe inflation\cite{guth} is well known as a promising theory to explain the origin of 
large scale structure of the Universe, a causal theory for the origin of primordial density 
fluctuations which may explain the observed density inhomogeneities and cosmic microwave 
fluctuations in the very early Universe, and to solve the  early universe pressing problems for the 
standard hot big bang theory\cite{tur}. For 
a resonable inflation model, the potential during inflation must be very flat in, at least, the 
direction of the inflaton. To construct a resonable inflation model, or the inflaton potential, all 
the known related astrophysics observations should be included. For a general tree-level hybrid 
inflation
 potential, which is not discussed fully so far for the quartic term, the parameters in it are shown 
how to be constrained via the astrophysics data observed and to be obtained to the expected accuracy 
by the soon lauched MAP and PLANCK satellite missions\cite{mp}, as well as the consistent cosmology 
requirements.
We find the effective inflaton mass parameter is in the TeV range, and the quartic term's 
self-coupling constant tiny, needs fine-tunning.
\end{abstract}

\vskip .3in



\end{titlepage}

\section{introduction}

For the past nearly twenty years, two major theories, inflation and topological defects to the early 
universe pressing problems\cite{tur} seem to have stood the test of time and one current goal is to 
determine which if any best fits the increasingly accumulated astrophysics data, especially the 
COBE's cosmic microwave background anisotropy detections as  well as the soon lauched more accuracy 
MAP and PLANCK observations\cite{mp}, and more reasonably interprete the origin of the Universe 
large scale structures. They usually are regarded as mutually exclusive theories in that defects 
formed before a period of inflation would rapidly be diluted to such a degree during the 
inflationary era as to make them of little interest to cosmology. But in some inflation models 
inspired from particle physics considerations the formation of topological defects can be naturally 
obtained at the end of the inflationary period\cite{lyth1}. On the other side, as is widely 
supposed, the initial conditions for the successful hot big bang are set by inflation, and then an 
adiabatic, Gaussian and more or less scale invariant density perturbation spectrum at horizon entry 
is predicted\cite{lyth}. Such a perturbation is generated by the vacuum fluctuation during inflation 
so the dazzling prospect of a window on the fundamental interactions on scales approaching the 
Planck energy appears. The studying of inflation paradigm will help us to understand basic physics 
laws to the possibly highest scale in the Nature.

The inflationary Universe scenario\cite{LL2} has the universe undergoing a period of accelerated 
expansion, the effect originally being to dilute monopoles (and any other defect formed before this 
period) outside of the observable universe, thereby dramatically reducing their density to below the 
observable limits. In a homogeneous, isotropic Universe with a flat Friedmann-Robertson-Walker (FRW) 
metric described by a scale factor a(t), the acceleration is given via 
\be
\ddot{a}=-a(4\pi G/3)(\rho+3p) 
\ee
where 
$\rho$ is the energy density and p the pressure. Usually the energy density which drives inflation 
is identified with a scalar potential energy density  that is positive, and flat enough to result in 
an effective equation of state  
\be
\rho\approx-p\approx V(\phi)
\ee
satisfying the acceleration condition $\ddot{a}>0$. 

The scalar potential is associated with a scalar field known as the so-called inflaton. During the 
inflationary period, the inflaton potential is fairly flat in the direction the field evolves, 
dominating the ennergy density of the universe for a finite period of time. Over the period it 
evolves slowly towards a minimum of the potential either through a classical roll over or through a 
quantum mechanics tunnelling transition. Inflation then ends when the inflaton starts to execute 
decaying oscillations around its own vacuum value, and the hot Big Bang (reheating) ensues when the 
vacuum value has been achieved and the decay products have thermalised. Over the past  decades there 
have been lots of inflation models constructed and there shall be certainly more with the coming of 
MAP and PLANCK satellites missions. With our knowledge so far we understand that any reasonable 
inflation model should satisfy at least that COBE normalization, cosmology observations constraint 
to the spectral index and adequate e-folding inflation for consistence requirements\cite{LL2}. In 
this line, the running-mass models of inflation without quartic term is studied\cite{mic} and we 
will discuss a general tree-level inflation model, especially the constraint to the quartic term 
self-coupling constant\cite{men}.        
 
This paper is arranged as following. In next section we give a general comments on the properties of 
inflaton potential, which must satisfy the COBE normalization condition besides the flatness 
conditions. In section three we examine detailly a tree-level hybrid inflation potential model, 
which can be regarded as a generalization of previously fully discussed some inflation 
models\cite{lind}. We use the slow-roll approximation to derive an analytic expression for the 
e-folds number N between a given epoch and the end of slow-roll inflation, and derive the spectral 
index of the spectrum of the curvature perturbation of this model.  Confronted them with the COBE 
measurement of the spectrum on large scales ( the normalization), the required e-folds number N and 
the observational constraint  on the spectrum index over the whole range of cosmological scales, we 
give the coupling constant of the inflaton model 
an allowed region, to specify  its parameter space by reducing its two free parameters to one .  
Finally we give a discussion and conclusion. 

\section{general considerations on Inflation model} 
Cosmological inflation has been regarded as the most elegant solution 
to the horizon and flatness problems of the standard Big Bang 
universe\cite{LL2}. If considering the later stage thermal inflation\cite{mod} it also beautifully 
solve the moduli problems\cite{lisa}. Even though it explains successfully why the current 
Universe appears so homogenous and flat in a very natural manner, it 
has been difficult to construct a model of inflation without a small 
parameter(the fine-tunning problem).  The key point is to have a 
resonable scalar field potential either from a more underlying 
 gravity theory like effective superstring thery or from a more fundmental particle physics theory 
such as supergravity or the so-called M-theory\cite{dav}. In fact
to any case, one needs at least a scalar field component(inflaton) that rolls 
down the potential very slowly with enough e-folds number to successfully generate a viable
inflationary scenario.
This requires the potential to be 
almost flat in the direction of the inflaton. There are lots of inflation models constructed so far. 
 If gravitation wave contribution is negligible, at least for the present situation, the curvature 
perturbation
spectrum index is the most powerful discriminator to inflation models.	In this section 
we discuss the general properties of an inflaton potential with the astrophysics considerations.
In the effective  slow-roll inflation scheme as physics requirments, the general inflaton potential 
$V(\phi)$
must satisfy the flatness conditions 
\be
\epsilon\ll 1
\ee
and 
\be
|\eta|\ll 1
\ee
where the notations\cite{LL2}
\bea
\epsilon &\equiv & \frac12 \mpl^2 (V'/V)^2 \\
\eta &\equiv & \mpl^2 V''/V
\eea
the prime indicates differiential with respect to $\phi$ and $\mpl = (8\pi G)^{-1/2} = 2.4\times 
10^{18}\GeV$ is the reduced
Planck mass(scale).  When these are satisfied, the time dependence of the
inflaton $\phi$ is generally given by the slow-roll expression
\be
3H\dot\phi = -V', 
\ee
where the quantity
\be
H\simeq \sqrt{\frac{1}{3}\mpl^{-2} V}
\ee 
is the Hubble parameter during inflation.
On a given scale, the spectrum of the primordial curvature perturbation,
thought to be the origin of structure in the Universe,
is given by
\be
\delta_H^2 (k) = \frac1{150\pi^2 \mpl^4} \frac V \epsilon
\ee
The right hand side is evaluated when the relevant scale
$k$ leaves the horizon. On large scales, the COBE observation
of the Cosmic Microwave Background(CMB) anisotropy corresponds to
\be
V^{1/4}/\epsilon^{1/4} = .027\mpl = 6.7 \times 10^{16}\GeV
\label{cobe}
\ee

The spectral index of the primordial curvature perturbation
is given by
\be
n-1 = 2\eta - 6\epsilon 
\ee
A perfectly scale-independent spectrum would correspond to $n=1$, 
and observation already demands 
\be
|n-1| < 0.2
\ee
Thus $\epsilon$ 
and $\eta$ have to be $\lsim 0.1$ (barring a cancellation)
and this constraint will get much tighter
if future observations, such as the near future MAP and PLANCK satellite experiment missions,
move $n$ closer to 1. Many models of inflation
predict that this will be the case, some giving a value of $n$ 
completely indistinguishable from 1.

Usually, $\phi$ is supposed to be charged under at least a
$Z_2$ symmetry, that is there is no change for the system under 
\be
\phi\to-\phi
\ee
which is unbroken during inflation.
Then $V'=0$ at the origin, and inflation typically takes place near the 
origin. As a result $\epsilon$ negligible compared with $\eta$, and
\be
n-1 = 2\eta\equiv 2\mpl^2 V''/V
\ee
We assume that this is the case as in most inflation models, in 
what follows. If it is not, the
inflation  model-building, as in the slow roll approximation, is even more tricky. 
Another point should be clear that the possibly later stage thermal inflation only lasts a few 
e-foldings and at a lower energy scale around the electroweak symmetry breaking scale, which affects 
less the extremally successful Hot 
Big Bang Nucleo-synthesis which is at the MeV scale. It takes place while a lighter scalar field 
(with mass around 100GeV) with nonzero vacuum expectation value, is trapped by thermal effects in 
the false vacuum at the lighter scalar field value as zero. More components inflation model only 
make physics picture complex. In this paper we consider two scalar fields one of which is the 
Higgs-like scalar field to make sure the graceful exit after the inflation end, and contribute to 
the vacuum energy expectation value when taking a critical value; another explicit one is the 
inflaton which drives the cosmic inflation (exponential) expansion.   

\section{A tree-level hybrid inflation  model}
In this section we will present the allowed parameter regions for a particular vaccum-dominated 
potential that are not  given out before. A similar form potential with only different signs appears 
in a Supersymmetry particle physics model\cite{riot}.
The focused vacuum dominated potential we consider has the usual form  with dimension four for the 
sake of renormalizability in mind  
\be
V=b{(M^2-h^2)^2}/2+m^2\phi^2/2 +\lambda\phi^4/4,.
\ee
where b is a coupling constant, m is the h expectation-value dependent mass parameter for the 
inflaton
and h is the Higgs-like scalar field for the graceful exit in the new inflation scenario\cite{lind}.
If the expectation value of h equals M during evolution
the model turns out to be the general chaotic inflationary potential\cite{ld};
if not then the usual hybrid inflation model with the
resulting $V_0$ as the dominated vacuum\cite{mod}.
\be 
V=V_0+m^2\phi^2/2 +\lambda\phi^4/4\label{v}\,.
\ee 
with all parameters positive in it, which can be regarded as a generalization of previously fully 
discussed inflation models\cite{lind}. Due
to symmetry considerations we discard the cubic term. Higher order
inflaton terms may appear in some Susy particle physics effective
models\cite{LL2,susy,dine}. There are two particular limits of vacuum energy inflation, according to 
whether the energy density is
dominated by the vacuum energy density or by the inflaton energy
density. We assume the former in our case as preference in the slow-roll approximation \cite{lyth}.
With the COBE normalization $ V^{1/4}/\epsilon^{1/4} = .027\mpl$ and
$\epsilon \ \equiv\  \frac12 \mpl^2 (V'/V)^2 $, in our case the false
vacuum energy density \be V_0\approx 0.3^4 \mpl^2
(m^2\phi_1+\lambda\phi_1^3)^{2/3}/2^{1/3} \,.
 \ee
 where $\phi_1$ is
the inflaton  value when COBE scale leaves the horizon.
To reduce free parameters number we define 
\be
y_1=m^2/\phi_1^2 . 
\ee
By cosmology observations to the power spectral index
constraint 
\be
\label{aa}
|n-1|/2=\eta= \mpl^2 V''/V<0.1
\ee
and if we take the nowadays
observation value upper limit 0.1 as a potentially changing parameter x, we have 
\be \eta=\mpl^2(m^2+3\lambda\phi^2)/V_0 < x\,.\ee
Taking the potential form into equation (\ref{aa}) and using our definition for $y_1$ we can define 
a function of $y_1$ as 
\be f(y_1)=\frac{y_1+3\lambda}{(y_1+\lambda)^{2/3}} < x  0.3^4/2^{1/3}
\label{7}\,.\ee
In this expression, the parameter $x$ as an observation input runs from nowadays 0.1 to the hopfully 
0.01 by the near future planned MAP and PLANK satellite missions. Easily we find
\be
df/dy_{1}=(y_{1}-3\lambda)(y_{1}+\lambda)^{-5/3}/3
\ee
For $y_{1}>3\lambda$ the $df/dy_{1}>0$, while for $y_{1}<3\lambda$ the $df/dy_{1}<0$, so that gives
\be
f_{min}(3\lambda)= 3(\lambda/2)^{-1/3}
\ee
which corresponds to, with relation (\ref{7}) into account 
\be
\lambda_{max}<x^{3} 1.97\times10^{-8}
\ee
With obviously 
\be
y_1+\lambda < y_1+3\lambda
\ee
and the relation (\ref{7}), directly we have
\be y_1+\lambda < x^3 0.027^4/2\,.\ee 
which together give the allowed parameters regions for quartic self coupling constant $\lambda$ and 
the reduced mass parameter with various observed or to be obtanined parameter x values as inputs.
Take today's upper limit $x=0.1$ we find the quartic self-coupling
constant $O(10^{-11})$, far too small. For the reduced mass parameter
it's normal since inflation starts at the inflaton field value around
0.1$M_{pl}$, which gives the inflaton's effective mass about 1000GeV.

With the certain e-folds number constraint for overcoming horizon and flatness problems 
\be N=|\int_{\phi_1} ^{\phi_c}\frac{ V}{V'}{ d\phi}|/\mpl^2\,.\ee
where $\phi_c$ is the inflaton at the end of inflation. Insert the potential form and the COBE 
normlization 
 we get another relation for the reduced mass parameter  defined as 
\be y_c=m^2/\phi_c^2=(\lambda+y_1)exp(Ny_1 10^3/3.6\times(\lambda+y_1)^{-2/3})-\lambda\,.\ee 
which limits the allowed parameters with astrophysics required e-folding numbers $N$ and the 
spectrum index x as today's observations required values\cite{lyt} as inputs. We can see the limit 
cases are consistent with the previous results by numerical calculations\cite{men} with 
figures\cite{lind}. The self-coupling constant is very tiny and
the fine-tunning problem appears.(We also can get directly from the above relation curves of the 
reduced parameter $y_c$ with parameter $\lambda$). It is clear that the reduced parameter $y_c$ is 
approximatly linear to $\lambda$ when the expenatial value is around 1.
 
Taking the power spectral index constraint relation (\ref{aa}) and the relation (\ref{7}) with the 
above equation into account we can get a constraint relation,  approximatedly allowed regions to 
reduced parameters $y_c$ and $\lambda$ as
\be
y_c+\lambda<(y_1+\lambda)exp(1000 N/3.608\times (y_1+\lambda)^{1/3})<2.657
\times10^{-7} x^3 exp(1.6518Nx)\ ,
\ee 
By 
which the approximatedly allowed parameter regions for  $\lambda$ vs $y_c$ with x from 0.1 to future 
possibly 0.01 is straightforward worked out.

 If put the power spectral index expression and the e-folding expression together we find 
\be |n-1|=N^{-1}2^{1/6}(1+3\lambda/y_1)ln(1+\frac{y_c-y_1}{\lambda+y_1})\,.\ee
qualitively that roughly is \be |n-1|\propto N^{-1}\,.\ee which implies these two constraints have 
intrinsic connection. When $|n-1|<0.1$ as today's cosmological observations available, then 
$N\gtrsim 10$. If $|n-1|<0.01$ as the soon satellite missions by MAP and PLANCK on design  hopefully 
to give, then the $N\gtrsim100$. 
This kind of generic character as expressed by relation (31) appears in a class of dynamical 
supersymmetry breaking particle physics models\cite{rio}.
\section{discussion and conclusion}
Models of inflation driven by a false vacuum formed by the Higgs-like scalar field are mainly 
different from true vacuum cases in their no zero false vacuum energy density, which are simple but 
also can reflect the astrophysics obervations. We discuss a regular tree-level hybrid inflation 
model, whose special case is the general chaotic inflation model(not a toy model\cite{la}) here to 
show how to constraint its parameters when confronting data and cosmology consistence requirements, 
and give several new  parameter relations and the allowed regions, which can be used as a prototype 
model for the two planned Microwave Anisotropy Probe (MAP) and PLANCK satellite missions tests. The 
results we have obtained based on a reasonable assumpation that the spectral index constraint we 
concentrate on is naturally satisfying the flatness conditions. Otherwise the slow roll 
approximation is not appliable.

The origin of this tree-level hybrid inflation model or its more complicated extentions may arise 
from some kind of supersymmetry particle physics or supergravity models which is generally cosidered 
as the appropriate framework for a description of the fundamental interactions at higher energy 
scale, and in particular for the description of their scalar interaction potential in the D-term and 
the gauge coupling F-Term\cite{king}. Here we only study the essence of them in order to get more 
viable inflation models from supersymmetry particle physics and supergravity as well as superstring 
(M-)theories. No matter what kind of theoretical model to be built it must satisfy at least the 
above observations, especially the spectral index constraint at x=0.01. The parameter space then in 
our case is very tiny that asks us to build the inflation models from a more natural way to avoid 
the fine-tunning problem, which is a chanlenge facing us.  Within next few years with the rapidly 
increasings in the variety and more accuracy of cosmological observation data, like the measurements 
of temperature anisotropies in cosmic microwave background at the accuracy expected from MAP and 
PLANCK soon, it is possible for us to discriminate among inflation models. 

As a flood of high-quality cosmological data is coming from experiments in outerspace, on earth and 
underground, even in the sea or at the bottom of the sea we are really entering the observation    
constrainning theory times. There are mainly six projects on the way. 
Here we only discuss the most related one to our topic, inflation model building:

The CMB Map of the Universe. COBE mapped the CMB with an angular resolution of around $10^0$; two 
new satellite missions, NASA's MAP(lauch in 2000) and ESA's PLANCK Surveyor (lauch 2007), will map 
CMB with 100 times better resolution ($0.1^0$) and with a detail map of our Universe at 300,000 
years. From these maps of the Universe as it existed at a simpler time, long before the first stars 
and galaxies, will come a gold mine of information: a definitive measurement of matter fraction of 
the Universe today;  a characterization of the primeval lumpiness and possible detection of the 
relic gravity waves from inflation as well as a determination of the Hubble constant to a precision 
of better than 0.05. Direct measurements of the expansion rate using standard candles, gravitational 
time delay, SZ imaging besides the precision CMB map will pin down the elusive Hubble constant once 
and for all. It is the fundamental parameter that sets the size- in time (the puzzling Universe age 
problem) and space- of the Universe. Its value is critical to testing the self consistency of the 
early Universe models. After all,  
the precision maps of the CMB that will be made are crucial to establishing the astroparticle 
physics inflation theory.

For the past two decades
the hot big bang model as been referred to as the standard cosmology- and for good reasons\cite{tn}. 
For just as long particle cosmologists have known there are fundamental questions that are not 
answered by the standard cosmology and point to a grander theory. The best candidate for that 
grander theory is the viable particle physics inflation theory plusing dark components (dark energy, 
hot and cold dark matter) as the Universe dominated contents. It holds that the Universe is flat, 
that slowly moving elemmentary particles left over from the earliest moments provide the cosmic 
infrastructure\cite{lm}, and that the primeval density inhomogeneities that seed all the structure 
arose from quantum fluctuations. There is now lots of prima facie evidence that supports the two 
basic tenets of this paradigm. An avalanche of high quality cosmological observations will soon make 
this case stronger or even will break it. Key questions remain to be answered; foremost among them 
are: a viable inflation model to be built, elucidation of the dark-energy component and the 
identification as well as detection of the cold dark matter particles. The next, at least, two 
decades are exciting times in Particle Physics Cosmology with the planned continuous astrophysics 
experiments going.

\vskip 1cm
\underline{\bf Acknowledgements}:
The author thanks Laura Covi, Robert Brandenberger, Ilia Gogoladze, Christopher Kolda, Xueqian Li, 
David Lyth, Leszek Roszkowski, Lewis Ryder, Goran Senjanovic, Roy Tung and Xinmin Zhang for many 
useful discussions on this topic during his working in Lancaster, UK and his visit to Abdum Salam 
ICTP, Italy. He also benefits greatly from comments to
his related work by R.Brandenberger, D.Lyth and X Zhang. 
This work is partially supported by grants
from the overseas research scholarship of China National Education Ministry and from Natural Science 
Foundation of China: the important project for theoretical physics key field research: Particle 
Physics Cosmology. 

\def\NPB#1#2#3{Nucl. Phys. {\bf B#1}, #3 (19#2)}
\def\PLB#1#2#3{Phys. Lett. {\bf B#1}, #3 (19#2) }
\def\PLBold#1#2#3{Phys. Lett. {\bf#1B} (19#2) #3}
\def\PRD#1#2#3{Phys. Rev. {\bf D#1}, #3 (19#2) }
\def\PRL#1#2#3{Phys. Rev. Lett. {\bf#1} (19#2) #3}
\def\PRT#1#2#3{Phys. Rep. {\bf#1} (19#2) #3}
\def\ARAA#1#2#3{Ann. Rev. Astron. Astrophys. {\bf#1} (19#2) #3}
\def\ARNP#1#2#3{Ann. Rev. Nucl. Part. Sci. {\bf#1} (19#2) #3}
\def\MPL#1#2#3{Mod. Phys. Lett. {\bf #1} (19#2) #3}
\def\ZPC#1#2#3{Zeit. f\"ur Physik {\bf C#1} (19#2) #3}
\def\APJ#1#2#3{Ap. J. {\bf #1} (19#2) #3}
\def\AP#1#2#3{{Ann. Phys. } {\bf #1} (19#2) #3}


\begin{thebibliography}{99}
\bibitem{tur} E.Kolb and M.Turner, The Early Universe (Addison-Wesley,Redwood City,CA 1990) and 
references therein; S.Weinberg,Gravitation and Cosmology, John Wiley(1972)
\bibitem{mic}L.Covi and D.Lyth, hep-th/9809562; L.Covi, D.Lyth and L.Roszkowski, hep-th/9809310.
\bibitem{ld} D.H.Lyth and A.Riotto, hep-ph/9807278 and  Phys.Rep. to appear; and references therein; 
G.Borner, The Early Universe, Springer-Verlag (1988)
\bibitem{dav}R.Brandenberger, astro-ph/9711106 and hep-ph/9407247
\bibitem{lyt}D.Lyth, hep-ph/9609431
\bibitem{rio} W.Kinney and A.Riotto, hep-ph/9704388
\bibitem{la}A.Lind and A.Riotto, hep-ph/9703209
\bibitem{lyth1}D.Lyth and A.Liddle, in Proc 2nd Paris Cosmology Colloquium, 1994, Paris (World 
Scientific Pub 1995)
\bibitem{lyth}E.Coplend,etc, in Proc 2nd Paris Cosmology Colloquium, 1994, Paris (World Scientific 
Pub 1995)
\bibitem{LL2} For a review of this material concerning inflation, see A.R.Liddle and D. H. Lyth, 
Phys. Rep. {\bf 231},
1 (1993); J.Lidsey, A.Liddle, E.Kolb, E.Copeland, T.Barriero and M.Abney, Rev.Mod.Phys.(1997) and 
astro-ph/9508078
\bibitem{guth}A.Guth, Phys.Rev.D23, 347(1981); A.Linde, Phys.Lett.B108,389(1982); A.Albrecht and 
P.Steinhardt, Phys.Rev.Lett.48,1220(1982); S.Hawking, Phys.Lett.B115,295(1982)
\bibitem{lisa}G.Dvali, Q.Shafi and R.Schaefer, Phys.Rev.Lett.73, 1886(1994); L.Randall, M.Soljacic 
and A.Guth, hep-ph/9601296 and Nucl.Phys.B472, 377(1996)
\bibitem{tn} M.Turner, astro-ph/9901168, Physica Scripta( in press astro-ph/9901109); M.Turner and 
J.Tyson, astro-ph/9901113 and references therein
\bibitem{mp} http://map.gsfc.nasa.gov/
\bibitem{men}Xinhe Meng, to appear in J.Phys.G, hep-ph/9809416, to appear in Comm. Theo.Phys., and 
hep-ph/9901624
\bibitem{mod}A.Liddle and D.Lyth, Cosmological Inflation and Large-Scale Structure, Oxford 
University Press (1999)
\bibitem{riot}M.Dine and A.Riotto, Phys.Rev.Lett.79,2632(1997)
\bibitem{susy} For reviews of supersymmetry and supergravity, see H.P.Nilles, Phys. Rep. {\bf 110}, 
1 (1984); H.E. Haber and G.L. Kane, Phys. Rept. {\bf 117}, 75 (1985) and  D. Bailin and A. Love, 
{\em Supersymmetric Gauge Field Theory  and String Theory}, IOP, Bristol (1994).
\bibitem{dine} A useful review of these questions is given by M. Dine, hep-th/9207045.
\bibitem{lind}A.Linde, Phys.Lett. B129,177(1983); Phys.Rev. D49,748(1994);Particle Physics and 
Inflation Cosmology, Harwood Academic, Switzerland(1990) and references therein
\bibitem{king} S.King and J.Sanderson, hep-ph/9707317 and in Cosmo-97
 First Int'l Workshop on Particle Physics and the Early Universe, Ambleside, England, Ed. Leszek 
Roszkowski, World Scientific(1998) and references therein
\bibitem{lm}J.Ellis, in Cosmo-97
 First Int'l Workshop on Particle Physics and the Early Universe, Ambleside, England, Ed. Leszek 
Roszkowski, World Scientific(1998) and references therein; R.Arnowitt and P.Nath, in Cosmo-97
 First Int'l Workshop on Particle Physics and the Early Universe, Ambleside, England, Ed. Leszek 
Roszkowski, World Scientific(1998) and references therein
; X.Li, Xinhe Meng, et al, Comm. Theor. Phys. 25(1996)342; ibid 26(1997)
\end{thebibliography}
\end{document}